# Convergent and divergent beam electron holography and reconstruction of adsorbates on free-standing two-dimensional crystals

*Tatiana Latychevskaia[1], Colin Robert Woods[2,3], Yi Bo Wang[2,3], Matthew Holwill[2,3], Eric Prestat[4], Sarah J. Haigh[2,4], Kostya S. Novoselov[2,3]*

[1]*Institute of Physics, Laboratory for ultrafast microscopy and electron scattering (LUMES), École Polytechnique Fédérale de Lausanne (EPFL) , CH-1015 Lausanne, Switzerland*

[2]*National Graphene Institute, University of Manchester, Oxford Road, Manchester, M13 9PL, UK*

[3]*School of Physics and Astronomy, University of Manchester, Oxford Road, Manchester, M13 9PL, UK*

[4]*School of Materials, University of Manchester, Oxford Road, Manchester, M13 9PL, UK*

**ABSTRACT**

Van der Waals heterostructures have been lately intensively studied because they offer a large variety of properties that can be controlled by selecting 2D materials and their sequence in the stack. The exact arrangement of the layers as well as the exact arrangement of the atoms within the layers, both are important for the properties of the resulting device. However, it is very difficult to control and characterize the exact position of the atoms and the layers in such heterostructures, in particular, along the vertical (z) dimension. Recently it has been demonstrated that convergent beam electron diffraction (CBED) allows quantitative three-dimensional mapping of atomic positions in three-dimensional materials from a single CBED pattern. In this study we investigate CBED in more detail by simulating and performing various CBED regimes, with convergent and divergent wavefronts, on a somewhat simplified system: a 2D monolayer crystal. In CBED, each CBED spot is in fact an in-line hologram of the sample, where in-line holography is known to exhibit high intensity contrast in detection of weak phase objects that are not detectable in conventional in-focus imaging mode. Adsorbates exhibit strong intensity contrast in zero and higher order CBED spots, whereas lattice deformation such as strain or rippling cause noticeable intensity contrast only in the first and higher order CBED spots. The individual CBED spots can be reconstructed as typical in-line holograms, and the resolution of 2.13 Å can be in principle achieved in the reconstructions. We provide simulated and experimental examples of CBED of a 2D monolayer crystal. The simulations show that individual CBED spots can be treated as in-line holograms and sample distributions such as adsorbates, can be reconstructed. Individual atoms can be reconstructed from a single CBED pattern provided the later exhibits high-order CBED spots. The experimental results were obtained in a transmission electron microscope (TEM) at 80 keV on free-standing monolayer hBN with some adsorbates. Examples of reconstructions obtained from experimental CBED patterns, at a resolution of 2.7 Å, are shown. CBED technique can be potentially interesting for imaging individual biological macromolecules, since it provides a relatively high resolution and does not require scanning procedure or multiple image acquisition and therefore allows minimizing the radiation damage.

Keywords: Graphene, Two-dimensional Materials, van der Waals Structures, Electron Holography, Convergent Beam Electron Diffraction

PACS: graphene, 81.05.ue; Electron holography in structure determination, 61.05.jp; Electron microscopy in structure determination, 68.37.-d; electron diffraction and scattering methods, 61.05.J-;

**MAIN TEXT**

# 1. Introduction

In convergent beam electron diffraction (CBED)[1], an electron wave of a wavelength much shorter than interatomic distances passes through the sample and its phase acquires a weak change caused by the interaction of the electron wave with the atomic potential of the sample. Because of the convergence of the incident wavefront, the intensity distribution of the scattered wave in the far field can be interpreted as acquired from a sample that is being illuminated at different angles. This in turn allows capturing three-dimensional information about atomic distributions in the sample. Recently, CBED imaging of two-dimensional crystals was demonstrated[2-4], where interpretation of CBED patterns is more straightforward than in the case of three-dimensional crystals[5-12]. For two-dimensional materials the interpretation of the intensity contrast in individual CBED spots is simpler than in the case of thick sample, because it directly maps the atomic distribution and deformations such as ripples. The sample position in a convergent electron beam can be relatively easy changed by moving the sample along the optical axis, which allows imaging the sample with convergent or divergent electron beam. A question often raised is whether imaging in either convergent or divergent electron beam is preferred. Below we answer this question by imaging the same sample in both imaging regimes. Another common question about CBED is the interpretation of a CBED pattern, since it combines properties of a diffraction pattern and of an in-line hologram (a defocused image) of the sample. The positions of the CBED spots are the same as positions of the diffraction peaks. The intensity distribution within a selected CBED spot is an in-line hologram (a defocused image) of the sample, it is formed when convergent/divergent wave illuminates the sample at an angle corresponding to the diffraction angle of that particular CBED spot. Therefore, the intensity distribution of each CBED spot is unique. It is also often discussed what resolution one can expect in the reconstruction obtained from a CBED pattern. On one hand, considering CBED pattern as a diffraction pattern, the resolution should be defined by the position of the highest detected diffraction peak. On the other hand, considering each CBED spot as an in-line hologram, the resolution should be defined by the Abbe criterion where the numerical aperture corresponds to the hologram (CBED spot) size[13]. In this work we address these issues by employing simulated and experimental CBED patterns.

A 2D monolayer crystal is a perfect test model to explain the intensity distributions formed in the zero and higher-order CBED spots. A general scheme of convergent beam electron diffraction experimental arrangement is shown in Fig. 1. Depending on the $z$-position of the sample (sample height), $\Delta f < 0$ or $\Delta f > 0$, the sample is illuminated with convergent or divergent incident wavefront, respectively, which corresponds to underfocus/overfocus imaging regime. The diameter of the imaged sample area linearly depends on the defocus distance as $|\Delta f| \cdot 2 \tan \alpha$, where $\alpha$ is the convergence semi angle as sketched in Fig. 1.

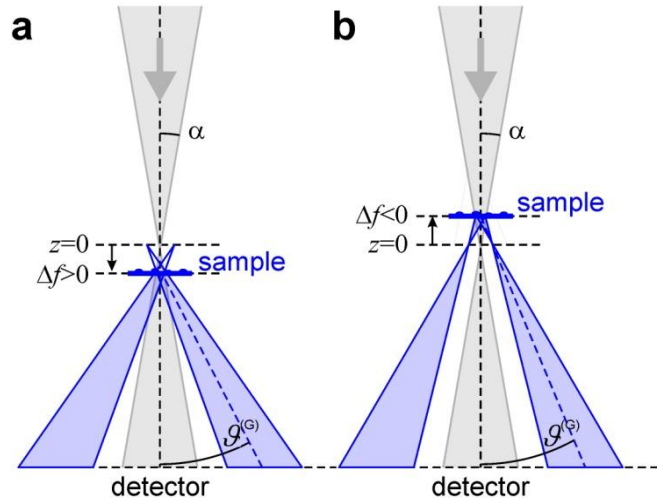

Fig. 1. Convergent beam electron diffraction experimental arrangement. (a) The sample (hBN) is positioned after the focus of the convergent beam, which corresponds to overfocus imaging conditions CBED, the sample height is $\Delta f > 0$. (b) The sample is positioned before the focus of the convergent beam, which corresponds to underfocus imaging conditions, the sample height is $\Delta f < 0$. $z = 0$ is the virtual source plane. $\alpha$ is the convergence semi angle. $\vartheta^{(G)}$ is the first-order diffraction angle.

# 2. Simulated results

## 2.2 Imaging lattice deformations

In this section we explain and compare the intensity contrast due to lattice deformations, such as out-of plane or in-plane ripples, when imaging in underfocus and overfocus regimes.

The distribution of the scattered wave in the far-field $U\left(\vec{R}\right)$ is given by

$$U\left(\vec{R}\right) \approx \int \frac{\exp\left(\pm ikr\right)}{r} t\left(\vec{r}\right) \frac{\exp\left(ik\left|\vec{r}-\vec{R}\right|\right)}{\left|\vec{r}-\vec{R}\right|} \mathrm{d}\vec{r} \propto \frac{\exp\left(ikR\right)}{R} \int \frac{\exp\left(\pm ikr\right)}{r} t\left(\vec{r}\right) \exp\left(-ik\frac{\vec{r}\vec{R}}{R}\right) \mathrm{d}\vec{r}, \quad (1)$$

where $\pm$ is for divergent/convergent wavefront, $t\left(\vec{r}\right)$ is the transmission function of the sample, $\vec{R}=\left(X,Y,Z\right)$ is the coordinate in the detector plane, $\vec{r}=\left(x,y,z\right)$ is the coordinate in the sample plane, and at $R\gg r$ the approximation $\left|\vec{r}-\vec{R}\right|\approx R-\frac{\vec{r}\vec{R}}{R}$ is applied. We introduce $K$-coordinates as

$$\vec{K}=\left(K_x,K_y,K_z\right)=k\frac{\vec{R}}{R}=\frac{2\pi}{\lambda R}\left(X,Y,Z\right), \quad \left|\vec{K}\right|=k=\frac{2\pi}{\lambda}, \quad K_z=\sqrt{K^2-K_x^2-K_y^2} \quad \text{and re-write:}$$

$$U\left(K_x,K_y\right)\approx \exp\left(ikR\right)\int \exp\left(\pm ikr\right)t\left(x,y,z\right)\exp\left[-i\left(K_x x+K_y y\right)\right]\exp\left(-izK_z\right)\mathrm{d}x\mathrm{d}y\mathrm{d}z. \quad (2)$$

### 2.2.1 Simulation procedure

Far-field wavefront distribution of the scattered wave was simulated by calculating the sum

$$U\left(K_x,K_y\right)=\sum_i L\left(\vec{r}_i\right)\psi_0\left(\vec{r}_i\right)\exp\left[-i\left(K_x x_i+K_y y_i\right)\right]\exp\left(-iz_i\sqrt{K^2-K_x^2-K_y^2}\right), \quad (3)$$

where $L\left(\vec{r}\right)$ is the distribution of atomic positions, $i$ runs through all the atoms in the lattice, $\vec{r}_i=\left(x_i,y_i,z_i\right)$ are the coordinates of the atoms, and $\psi_0\left(\vec{r}\right)$ is the incident wavefront distribution. $\psi_0\left(\vec{r}\right)$ is calculated by simulation diffraction of the spherical wavefront on a limiting aperture (second condenser aperture) positioned at a plane $\vec{r}_0$:

$$\psi_0\left(\vec{r}\right)=\iint a\left(\vec{r}_0\right)\frac{\exp\left(-ikr_0\right)}{r_0}\frac{\exp\left(ik\left|\vec{r}_0-\vec{r}\right|\right)}{\left|\vec{r}_0-\vec{r}\right|}\mathrm{d}\vec{r}_0, \quad (4)$$

where $a\left(\vec{r}_0\right)$ is the aperture function. CBED pattern was then calculated as $I\left(K_x,K_y\right)=\left|U\left(K_x,K_y\right)\right|^2$.

In the next subsections we derived the phase shifts introduced by in-plane and out-of-plane sample ripples, when imaging in both convergent and divergent modes. A corresponding sketch for atomic shifts is shown in Fig. 2.

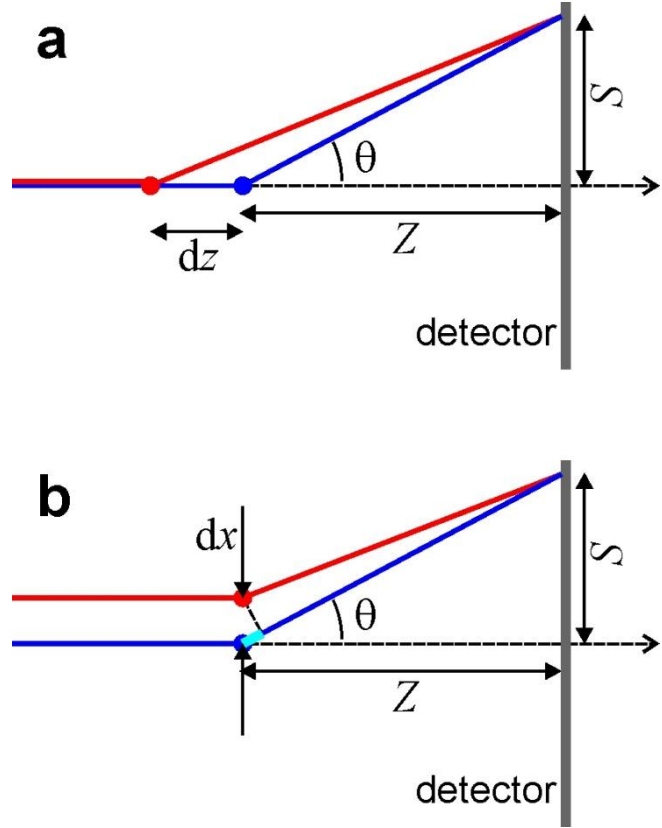

Fig. 2. Geometrical arrangement of scattering from two atoms positioned at different (a) $z$ and (b) $x$-distances. (a) The path difference is given by $l_{\text{blue}} - l_{\text{red}}$, where $l_{\text{blue}} = \mathrm{d}z + \sqrt{Z^2 + S^2} \approx \mathrm{d}z + \dfrac{Z}{\cos \vartheta}$ and $l_{\text{red}} = \sqrt{(\mathrm{d}z + Z)^2 + S^2} \approx \dfrac{Z}{\cos \vartheta} + \mathrm{d}z \cos \vartheta$, so that $l_{\text{blue}} - l_{\text{red}} \approx \mathrm{d}z (1 - \cos \vartheta)$. (b) The optical path difference (shown in cyan) is given by $\mathrm{d}x \sin \vartheta$.

### 2.2.2 Phase shift caused by an out-of-plane displacement, convergent mode

For convergent mode, $\Delta f < 0$, the wavefront scattered by an atom at $\vec{r} = (x, y, z)$ is given by Eq. 2:

$$U(K_x, K_y) \propto \exp(ikR)\exp(-ikr)\exp\left[-i\left(xK_x + yK_y\right)\right]\exp(-izK_z). \tag{5}$$

Wavefronts scattered by atoms positioned at $\vec{r}_1 = (0, 0, -|\Delta f|)$, $r_1 = |\Delta f|$ and $\vec{r}_2 = (0, 0, -|\Delta f| + \Delta z)$, $r_2 = |-|\Delta f| + \Delta z|$, are given by:

$$
\begin{aligned}
U_1(K_x, K_y) &\propto \exp(ikR)\exp\left(-ik|\Delta f|\right)\exp\left[iK_z|\Delta f|\right] \\
U_2(K_x, K_y) &\propto \exp(ikR)\exp\left[ik\left(-|\Delta f| + \Delta z\right)\right]\exp\left[-iK_z\left(-|\Delta f| + \Delta z\right)\right].
\end{aligned} \tag{6}
$$

The corresponding phases of the wavefronts are:

$$\varphi_1 = kR - k\left|\Delta f\right| + \Delta f K_z$$
$$\varphi_2 = kR + k\left(-\left|\Delta f\right| + \Delta z\right) - \left(-\left|\Delta f\right| + \Delta z\right)K_z,$$

(7)

and the phase difference is given by:

$$\Delta \varphi_z = \varphi_2 - \varphi_1 = \Delta z\left(k - K_z\right) = \Delta z \frac{2\pi}{\lambda}\left(1 - \cos\vartheta\right),$$

(8)

where we applied $K_z = \frac{2\pi}{\lambda}\cos\vartheta$. The sketch of the optical path difference is shown in Fig. 2a. The corresponding simulations are shown in Fig. 3.

### 2.2.3 Phase shift caused by an out-of-plane displacement, divergent mode

For divergent mode, $\Delta f > 0$, the wavefront scattered by an atom at $\vec{r} = (x, y, z)$ is given by Eq. 2:

$$U\left(K_x, K_y\right) \propto \exp\left(ikR\right)\exp\left(ikr\right)\exp\left[-i\left(xK_x + yK_y\right)\right]\exp\left(-izK_z\right).$$

(9)

Wavefronts scattered by atoms positioned at $\vec{r_1} = (0, 0, \Delta f)$, $r_1 = \Delta f$ and $\vec{r_2} = (0, 0, \Delta f + \Delta z)$,

$r_2 = \Delta f + \Delta z$ are given by:

$$U_1\left(K_x, K_y\right) \propto \exp\left(ikR\right)\exp\left(ik\Delta f\right)\exp\left[-iK_z\Delta f\right]$$
$$U_2\left(K_x, K_y\right) \propto \exp\left(ikR\right)\exp\left[ik\left(\Delta f + \Delta z\right)\right]\exp\left[-iK_z\left(\Delta f + \Delta z\right)\right].$$

(10)

The corresponding phases of the wavefronts are:

$$\varphi_1 = kR + k\Delta f - \Delta f K_z$$
$$\varphi_2 = kR + k\left(\Delta f + \Delta z\right) - \left(\Delta f + \Delta z\right)K_z,$$

(11)

and the phase difference is given by:

$$\Delta \varphi_z = \varphi_2 - \varphi_1 = \Delta z\left(k - K_z\right) = \Delta z \frac{2\pi}{\lambda}\left(1 - \cos\vartheta\right),$$

(12)

where we applied $K_z = \frac{2\pi}{\lambda}\cos\vartheta$. The sketch of the optical path difference is shown in Fig. 2a. The phase difference is the same as in the case of convergent probing wavefront. The corresponding simulations are shown in Fig. 3, the intensity inversion will be explained later.

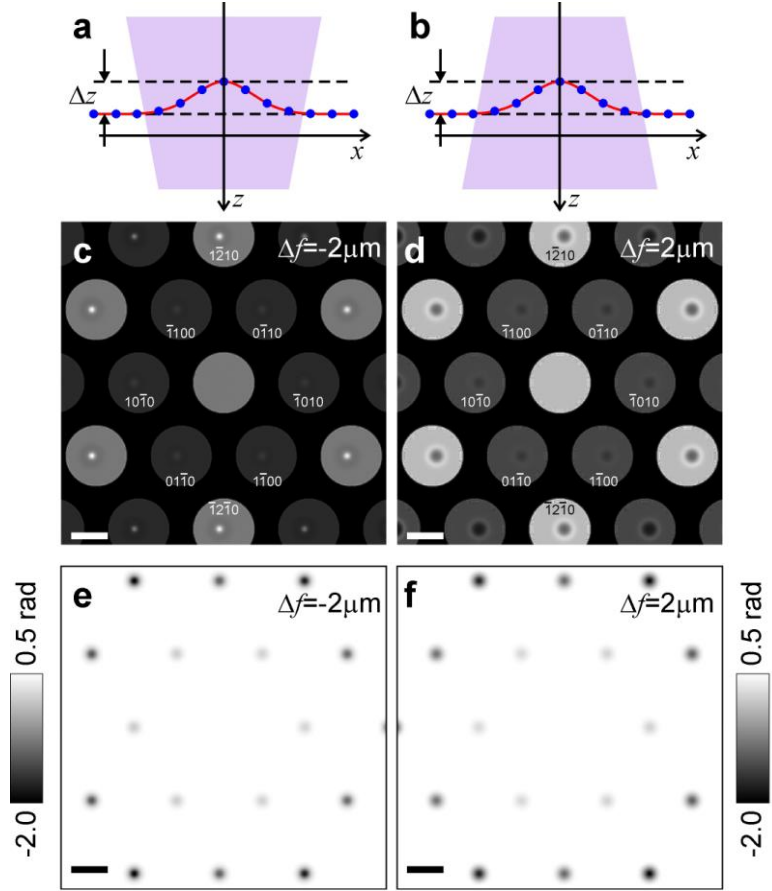

Fig. 3. CBED patterns of a graphene monolayer with a an out-of plane bubble simulated for $\Delta f$ = -2μm (underfocus) and $\Delta f$ = 2μm (overfocus). (a) and (b) Sketches of the side view of a graphene layer with atoms displaced out of plane due to the presence of a bubble. (c) and (d) the corresponding simulated CBED patterns. (e) and (f) phase shift introduced by the lattice deformation into the probing electron wave, calculated at the detector plane. For these simulations the bubble height is $\Delta z$ = 2 nm the imaged area is about 30 nm in diameter, the number of pixels is 512 × 512 and $\Delta K = 3.51 \times 10^7$ m$^{-1}$. The scale bars in (c) – (f) correspond to 2 nm$^{-1}$.

### 2.2.4 Phase shift caused by an in-plane displacement, convergent mode

For convergent mode, $\Delta f < 0$, the wavefront scattered by an atom at $\vec{r} = (x, y, z)$ is given by Eq. 2:

$$U\left(K_x, K_y\right) \propto \exp\left(ikR\right)\exp\left(-ikr\right)\exp\left[-i\left(xK_x + yK_y\right)\right]\exp\left(-izK_z\right). \qquad (13)$$

Wavefronts scattered by atoms positioned at $\vec{r_1} = \left(0, 0, -|\Delta f|\right)$ and $\vec{r_2} = \left(a + \Delta x, 0, -|\Delta f|\right)$, where $r_1 = |\Delta f|$ and $r_2 = \sqrt{\left(\Delta f\right)^2 + \left(a + \Delta x\right)^2} \approx |\Delta f|$, and $a$ is the lattice period, are given by:

$$U_1(K_x, K_y) \propto \exp(ikR)\exp(-ik|\Delta f|)\exp[iK_z|\Delta f|]$$
$$U_2(K_x, K_y) \propto \exp(ikR)\exp(-ik|\Delta f|)\exp[-iK_x(a+\Delta x)]\exp(iK_z|\Delta f|). \tag{14}$$

The corresponding phases of the wavefronts are:

$$\varphi_1(K_x, K_y) = kR - k|\Delta f| + K_z|\Delta f|$$
$$\varphi_2(K_x, K_y) = kR - k|\Delta f| - K_x(a+\Delta x) + K_z|\Delta f|, \tag{15}$$

and the phase difference is given by:

$$\Delta\varphi_x(K_x, K_y) = \varphi_2(K_x, K_y) - \varphi_1(K_x, K_y) = -K_x(a+\Delta x). \tag{16}$$

When $\Delta x = 0$, the phase shift $\Delta\varphi = 2\pi$, and we obtain:

$$\Delta\varphi_x(K_x, K_y) = -K_x^{(1)}a = 2\pi, \tag{17}$$

which corresponds to the position of the $n$-th-order diffraction peak

$$K_x^{(n)} = n\frac{2\pi}{a}. \tag{18}$$

The phase shift due to a lateral shift $\Delta x$ is given by

$$\Delta\varphi_x(K_x, K_y) = -K_x\Delta x = -\frac{2\pi}{\lambda}\Delta x \sin\vartheta \tag{19}$$

which is an odd function of $K_x$. Thus, for $\Delta x \neq 0$ there will be additional phase shift in opposite CBED spots, as for example in the spots $(\bar{1}010)$ and $(10\bar{1}0)$, and these phase shifts will be of opposite sign. The sketch of the optical path difference is shown in Fig. 2b. The corresponding simulations are shown in Fig. 4.

## 2.2.5 Phase shift caused by an in-plane displacement, divergent mode

For divergent mode, $\Delta f > 0$, the wavefront scattered by an atom at $\vec{r} = (x, y, z)$ is given by Eq. 2:

$$U(K_x, K_y) \propto \exp(ikR)\exp(ikr)\exp[-i(xK_x + yK_y)]\exp(-izK_z). \tag{20}$$

Wavefronts scattered by atoms positioned at $\vec{r_1} = (0, 0, \Delta f)$ and $\vec{r_2} = (a+\Delta x, 0, \Delta f)$, where $r_1 = \Delta f$ and $r_2 = \sqrt{(\Delta f)^2 + (a+\Delta x)^2} \approx \Delta f$, and $a$ is the lattice period, are given by:

$$U_1(K_x, K_y) \propto \exp(ikR)\exp(ik\Delta f)\exp[-iK_z\Delta f]$$
$$U_2(K_x, K_y) \propto \exp(ikR)\exp(ik\Delta f)\exp[-iK_x(a+\Delta x)]\exp(-iK_z\Delta f). \tag{21}$$

The corresponding phases of the wavefronts are:

$$\varphi_1(K_x, K_y) = kR + k\Delta f - K_z\Delta f$$
$$\varphi_2(K_x, K_y) = kR + k\Delta f - K_x(a+\Delta x) - K_z\Delta f, \tag{22}$$

and the phase difference is given by:

$$\Delta\varphi_x\left(K_x,K_y\right) = \varphi_2\left(K_x,K_y\right) - \varphi_1\left(K_x,K_y\right) = -K_x\left(a+\Delta x\right). \tag{23}$$

When $\Delta x = 0$, the phase shift $\Delta\varphi = 2\pi$, and we obtain:

$$\Delta\varphi_x\left(K_x,K_y\right) = -K_x^{(1)}a = 2\pi, \tag{24}$$

which corresponds to the position of the $n$-th-order diffraction peak

$$K_x^{(n)} = n\frac{2\pi}{a}. \tag{25}$$

The phase shift due to a lateral shift $\Delta x$ is given by

$$\Delta\varphi_x\left(K_x,K_y\right) = -K_x\Delta x = -\frac{2\pi}{\lambda}\Delta x \sin\vartheta, \tag{26}$$

which is an odd function of $K_x$. Thus, for $\Delta x \neq 0$ there will be additional phase shift in opposite CBED spots, as for example, $\left(\overline{1}010\right)$ and $\left(10\overline{1}0\right)$, and these phase shifts will be of opposite sign. The sketch of the optical path difference is shown in Fig. 2b. The corresponding simulations are shown in Fig. 4. Although the formulas for the phase difference is the same for convergent/divergent wavefront, the intensity distribution within a CBED spot is mirror-symmetrically flipped for convergent/divergent wavefront, as evident from Fig. 4.

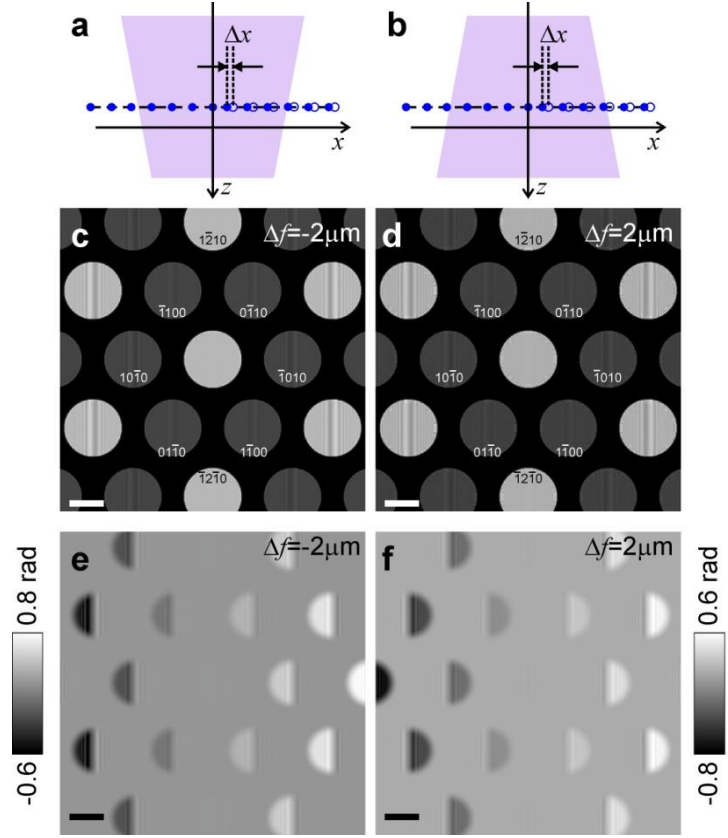

Fig. 4. Simulated CBED patterns of a graphene monolayer with in-plane atomic shifts, where the atoms positioned at $x>0$ are displaced by $\Delta x$=-10 pm. CBED patterns are simulated for $\Delta f$ = -2μm (underfocus) and $\Delta f$ = 2μm (overfocus). (a) and (b) sketches of the side view of a graphene layer with atoms displaced out of plane due to the presence of a bubble. (c) and (d) the corresponding simulated CBED patterns. (e) and (f) phase shift introduced by the lattice deformation into the probing electron wave, calculated at the detector plane. For these simulations the bubble height is $\Delta z$ = 2 nm the imaged area is about 30 nm in diameter, the number of pixels is 512 × 512 and $\Delta K = 3.51 \times 10^7$ m$^{-1}$. The scale bars in (c) – (f) correspond to 2 nm$^{-1}$.

## 2.3. Imaging adsorbates

In this section we will explain the intensity contrast formation caused by the presence of adsorbates such as patches of a second layer or individual molecules when imaged in underfocus and overfocus regimes. We will also demonstrate how adsorbates can be reconstructed from individual CBED spots and discuss the achievable resolution.

One might think that CBED imaging at $\Delta f < 0$ and $\Delta f > 0$ should provide the same results. This is correct except for when the imaged object is a weak phase object, in which case the CBED patterns acquired at $\Delta f < 0$ and $\Delta f > 0$ exhibit opposite contrast. This effect can be explained as follows. The interference pattern is given by $U_O U_R^* + U_O^* U_R$, where $U_O U_R^*$ is the object term and $U_O^* U_R$ is the twin image term[14, 15], where $U_R = \exp(ikR)$ and $U_O \approx \int o(\vec{r}) \exp\left(ik\left|\vec{r} - \vec{R}\right|\right) d\vec{r} \approx \exp(ikR) \int o(\vec{r}) \exp(-ik\vec{K}\vec{r}) d\vec{r}$. This gives $U_O U_R^* \approx \int o(\vec{r}) \exp(-i\vec{K}\vec{r}) d\vec{r}$ and $U_O^* U_R \approx \int o^*(\vec{r}) \exp(i\vec{K}\vec{r}) d\vec{r}$, respectively.

During reconstruction procedure, the object term $U_O U_R^*$ gives $u(\vec{r}) = \int o(\vec{r}) \exp(-i\vec{K}\vec{r}) d\vec{r} \int \exp(i\vec{K}\vec{r}\,') d\vec{K} = o(\vec{r}\,')$ and the twin term $U_O^* U_R$ gives $u(\vec{r}\,') = \int o^*(\vec{r}) \exp(i\vec{K}\vec{r}) d\vec{r} \int \exp(i\vec{K}\vec{r}\,') d\vec{K} = o^*(-\vec{r}\,')$. A schematics of the positions of the reconstructed object and its twin image is shown in Fig. 5, which illustrates that a complex-valued object $o(\vec{r})$ imaged in CBED at $\Delta f > 0$ (positioned at $z > 0$) and a complex-valued object $o^*(-\vec{r})$ imaged in CBED at $\Delta f < 0$ (positioned at $z < 0$) give rise to the same CBED pattern. The following happens when the same object is imaged in divergent/convergent modes. For a weak phase object the following approximation holds $o(\vec{r}) = \exp\left[i\Delta\varphi(\vec{r})\right] \approx 1 + i\Delta\varphi(\vec{r})$, and the interference pattern measured in CBED at $\Delta f > 0$ (object positioned at $z > 0$) is given by

$$\left(U_O U_R^* + U_O^* U_R\right)_{\Delta f > 0}(\vec{K}) \sim i \int \Delta\varphi(\vec{r}) \left[\exp(-i\vec{K}\vec{r}) - \exp(i\vec{K}\vec{r})\right] d\vec{r}.$$

Positioning the same object at $z < 0$, that is measuring it in CBED $\Delta f < 0$ mode gives an interference pattern described by

$$
\begin{aligned}
\left(U_O U_R^* + U_O^* U_R\right)_{\Delta f < 0}(\vec{K}) &\sim i \int \Delta\varphi(-\vec{r}) \left[\exp(-i\vec{K}\vec{r}) - \exp(i\vec{K}\vec{r})\right] d\vec{r} = \\
&-i \int \Delta\varphi(\vec{r}\,') \left[\exp(i\vec{K}\vec{r}\,') - \exp(-i\vec{K}\vec{r}\,')\right] d\vec{r}\,' = \\
&-\left(U_O U_R^* + U_O^* U_R\right)_{\Delta f > 0}(-\vec{K})
\end{aligned}
\tag{27}
$$

which is a spatially centro-symmetrical-flipped distribution, and when added to constant intensity term $|U_R|^2$ it creates to an inverted intensity distribution. This effect of inversion of the contrast and centro-symmetrical flipping of the distributions within CBED spots is evident in the simulations shown in Fig. 3, and will be also demonstrated in the experimental CBED patterns in the following sections.

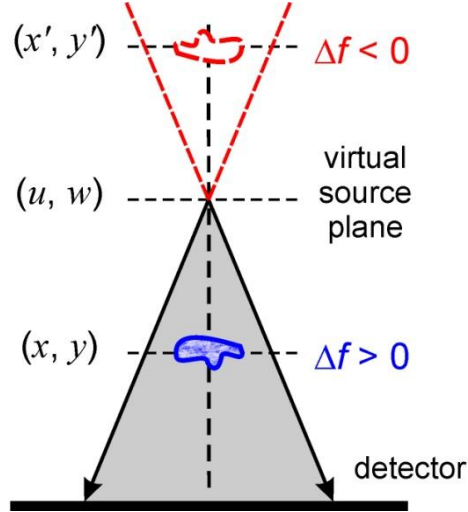

Fig. 5. Convergent beam electron diffraction experimental arrangement where it is shown that an object imaged in $\Delta f > 0$ arrangement is the same as complex-conjugated object imaged in $\Delta f < 0$ arrangement.

Monolayer can be described by transmission function $t(x,y) = \exp\left[i\sigma V_z(x,y)\right]$, where $\sigma = \dfrac{2\pi me\lambda}{h^2}$, $V_z(x,y)$ is the specimen projected potential, $m$ is the relativistic mass of electron, $e$ is an elementary charge, and $h$ is Planck's constant. For example, for graphene, at a radius of 0.1 Å the projected atomic potential of a carbon atom is 0.215 kV·Å and the interaction parameter $\sigma$ is 1.008 radian/kV·Å at a beam energy of 80 keV. This means that a single carbon atom will produce a total phase shift of 0.22 radians,[16] and graphene monolayer can be considered as a weak phase sample.

As an example, we simulated and reconstructed CBED pattern of monolayer graphene (G) with a patch (P) on top, the results are presented in Figure 6. The far-field distribution of the scattered wave was simulated by calculating the following integral transformation[17, 18]:

$$U\left(K_x, K_y\right) = -\frac{i}{\lambda} \iint \psi_0(x,y) t(x,y) \exp\left[-i\left(K_x x + K_y y\right)\right] \exp\left(-iz\sqrt{K^2 - K_x^2 - K_y^2}\right) \mathrm{d}x\mathrm{d}y,$$

(28)

where $\psi_0(x,y)$ is the incident wavefront calculated as defined by Eq. 4, $t(x,y) \approx \exp\left\{i\sigma\left[V_{G,z}(x,y) + V_{P,z}(x,y)\right]\right\}$, $V_{G,z}(x,y)$ is the total projected potential of the graphene molnolayer and $V_{P,z}(x,y)$ is the total projected potential of the patch. The total specimen projected potential $V_{i,z}(x,y)$, $i$ = G, P was calculated as a sum $V_{i,z}(\vec{r}) = \sum_{j=1}^{N} v_{i,z}(\vec{r} - \vec{r}_j)$, where $\vec{r}_j$ is coordinate of $j$-th atom and sum is performed over all atoms in the sample; $v_{i,z}(\vec{r})$ is projected potential of a single carbon atom, it is calculated as explained in Ref.[16]. Figure 6a and b show the simulated phase distribution of the transmission function of the sample. The total specimen potential was calculated by shifting the projected potential of a single atom to the position of each atom in the sample and summing the shifted potentials up (alternatively, it can be simulated via convolution). The complex-valued wavefront $U(K_x, K_y)$ was calculated according to Eq. 28 by applying single FFT. The CBED pattern was then calculated as $I(K_x, K_y) = \left| U(K_x, K_y) \right|^2$, shown in Fig. 6c. From the distributions of the individual CBED spots it is apparent that the patch acts as a weak phase object and creates an intensity contrast that is equal for all order CBED spots. This is a different situation than in the case of lattice deformations, where the contrast increases for higher order CBED spots. Thus, a weak phase objects and lattice deformations can be clearly separated from each other by simple comparing the contrast in high-order CBED spots.

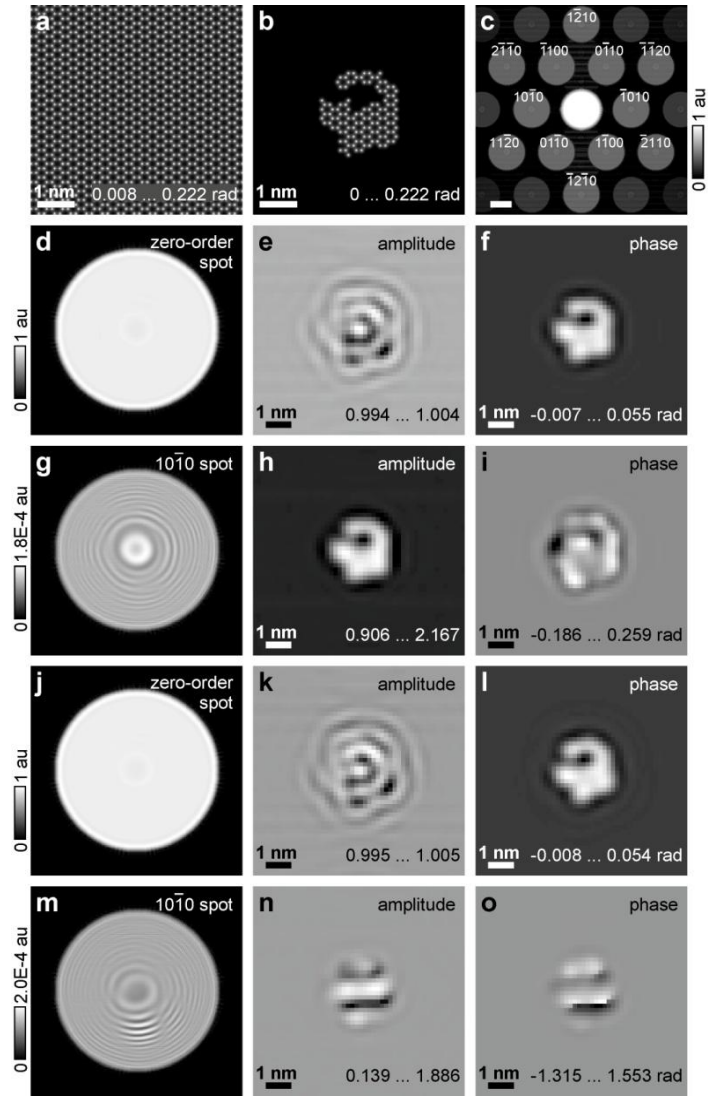

Fig. 6. Reconstruction of CBED spots as an in-line holograms from simulated CBED pattern of graphene with a patch of graphene on top.

(a) Phase distribution of the simulated transmission function of a graphene monolayer.

(b) Phase distribution of the simulated transmission function of a patch of the second graphene layer residing on top of the first layer.

(c) Simulated CBED pattern, $\Delta f = -2$ µm, imaged area is about 40 nm in diameter, electron energy 80 keV, intensity shown in logarithmic scale, scalebar is 2 nm$^{-1}$.

(d) Zero-order CBED spot and reconstructed amplitude (e) and phase (f) distributions.

(g) First-order CBED spot and reconstructed amplitude (h) and phase (i) distributions.

(j) The patch is rotated by 10°: Zero-order CBED spot and the corresponding reconstructed amplitude (k) and phase (l) distributions.

(m) The patch is rotated by 10°: First-order CBED spot and the corresponding reconstructed amplitude (n) and phase (o) distributions.

### 2.3.1 Reconstruction of individual CBED spots

Next, the zero-order and a first-order CBED spots were reconstructed individually as in-line holograms. It was assumed that the wavefront phase distribution in the detector plane is known. This was done to create an ideal situation (without twin image term), and to see what values of phase and amplitude are reconstructed in such an ideal case. The reconstructions were obtained by solving the following integral transformation[17-19]:

$$u(x,y,z) = \frac{i}{\lambda} \iint U(K_x, K_y) \exp\left[i(K_x x + K_y y)\right] \exp\left(iz\sqrt{K^2 - K_x^2 - K_y^2}\right) dK_x dK_y, \quad (29)$$

which was calculated by applying two FFTs as explained in Ref.[18]. The amplitude and the phase distributions were extracted from the reconstructed distribution as $|u(x,y,z)|$ and $\text{Arg}\left[u(x,y,z)\right]$, respectively.

To interpret the contrast formation in an individual CBED spot and its reconstruction we apply the following consideration. For a graphene monolayer with adsorbates residing on top, weak phase object approximation applies and the transmission function can be written as:

$$t(x,y) \approx 1 + i\sigma V_{G,z}(x,y) + i\sigma V_{P,z}(x,y). \quad (30)$$

The wavefront in the far-field can be represented as a sum of three terms: $U(\vec{R}) = U_0(\vec{R}) + U_1(\vec{R}) + U_2(\vec{R})$ that correspond to the three terms in the transmission function as described by Eq. 30. The resulting intensity distribution is given by $|U(\vec{R})|^2$. Each CBED spot can be interpreted as an in-line hologram which is formed by interference between a reference and an object waves. In the zero-order CBED spot the reference wave is provided by the non-diffracted wave (term "1" in Eq. 30) and the object wave is provided by the diffracted wave (term "$i\sigma V_{G,z}(x,y) + i\sigma V_{P,z}(x,y)$" in Eq. 30). In the higher-order CBED spots there is only diffracted wave (no term "1"), so that the reference wave is provided by the wave scattered on the graphene layer (term "$i\sigma V_{G,z}(x,y)$") and the object wave is provided by the wave diffracted on the adsorbates layer (term "$i\sigma V_{P,z}(x,y)$").

Before reconstruction the selected CBED spot was multiplied with apodization function which blurs the sharp edges and set the values around the spot to a constant value, this helps to avoid effect of diffraction on edges (concentric fringes) superimposed onto the reconstruction[18].

The reconstruction obtained from the zero-order CBED spot is shown in Fig. 6d − f. Here the contrast of the reconstructed amplitude is almost zero, while the reconstructed phase distribution resembles the patch shape and quantitatively varies from 0 to 0.06 radian, which is less than a phase shift expected from a single carbon atom (0.217 radian). The discrepancy can be explained by the fact that the obtained low-resolution reconstruction exhibits phase distribution that is averaged over the patch.

The reconstruction obtained from the first-order CBED spot is shown in Fig. 6g − i. As already mentioned, in the higher-order CBED spots the graphene layer provides the reference wave (term "$i\sigma V_{G,z}(x, y)$") and the adsorbates provide the object wave (term "$i\sigma V_{P,z}(x, y)$"). Therefore, the interference is provided by $V_{G,z}(x, y) + V_{P,z}(x, y)$ term that is a real-valued distribution that is expected to be reconstructed as amplitude. The reconstructions shown in Fig. 6 − i demonstrate that indeed the reconstructed amplitude distribution resembles the patch shape, while the phase distribution is close to zero. The reconstructed amplitude can be attributed to $V_{G,z}(x, y)$ in the areas surrounding the patch and to the sum $V_{G,z}(x, y) + V_{P,z}(x, y)$ in the patch region. Because in the simulations we assumed the same projected potential of atoms in the graphene layer and in the patch, the reconstructed amplitude values are 1 au and 2 a.u. in these areas, correspondingly.

***Rotated patch.*** To mimic situation when the deposited object consists of atoms whose positions do not exactly match the atomic positions of the supporting lattice, as for example in the case of mismatching lattices (Moire patterns) or a biological macromolecule, we simulated a patch whose lattice was rotated relatively to the support lattice by 10°, Fig. 6j − o. Note that the intensity distribution in the first- and higher CBED spots changed: the in-line hologram formed by the patch remains at the same position within CBED spot (centered), but the fringe pattern becomes asymmetrical, compare Fig. 6g and m.

The reconstructions obtained from the zero-order CBED spot are shown in Fig. 6j − l, they are almost the same as obtained for the not-rotated patch: amplitude and phase distributions are correctly reconstructed, compare Fig. 6j − l and Fig. 6e − f. The amplitude and phase distributions reconstructed from the first-order CBED spot, shown in Fig. 6m − o, cannot be easily interpreted. The reason is that the interference pattern in a CBED spot is formed by two wavefronts originating from two virtual sources. When the period and the orientation of the lattices of the graphene support and the patch are matching, the two virtual sources share almost the same position. When patch is rotated, the position of the corresponding virtual source is also rotated from its original position, this creates an additional linear phase shift in the CBED spot plane which is then superimposed onto the reconstruction. In principle, when the exact rotation of the patch is known, the linear phase shift can be compensated and correct phase distribution of the patch can be reconstructed. This is however only possible when the lattice period of the patch is the same as of the supporting lattice, which is not true in a general case of a non-periodical macromolecule. In the case of non-periodical object, the wavefront scattered by the object does not exhibit the same diffraction spots as the supporting lattice. Thus, a non-periodical object residing on a crystalline support can be reconstructed only from either the zero-order CBED spot (as we showed in Fig. 6d − f and j − l) or the entire CBED pattern (as we show below).

### 2.3.2 Reconstruction of entire CBED pattern

The reconstruction obtained from individual CBED spots exhibit resolution limited by the CBED spot size and a higher resolution can be in principle achieved when reconstructing the entire CBED pattern at once. The intensity distribution of a CBED pattern can be written in form of "holographic" equation:

$$\mathrm{CBED} = \left|U_\mathrm{G}\right|^2 + \left|U_\mathrm{P}\right|^2 + U_\mathrm{G}^* U_\mathrm{P} + U_\mathrm{G} U_\mathrm{P}^*, \qquad (31)$$

where $U_\mathrm{G}$ and $U_\mathrm{P}$ are the far-field distributions of the waves scattered by graphene monolayer and the patch, correspondingly. Here $U_\mathrm{G}$ can be considered as a "reference" wave, whose distribution can be simulated for a perfect (no defects or ripples) graphene monolayer. Square root of the CBED pattern distribution multiplied with the phase distribution of $U_\mathrm{G}$ provides a good estimation of $U_\mathrm{P}$:

$$\sqrt{\mathrm{CBED}} \approx \left|U_\mathrm{G}\right| + \frac{U_\mathrm{G}^* U_1 + U_\mathrm{G} U_\mathrm{P}^*}{2\left|U_\mathrm{G}\right|} \qquad (32)$$

$$\frac{U_\mathrm{G}}{\left|U_\mathrm{G}\right|}\sqrt{\mathrm{CBED}} \approx U_\mathrm{G} + \frac{U_P}{2}. \qquad (33)$$

By propagating backward the wavefront from the detector plane to the object plane we obtain the patch reconstruction as a weak phase object (Eq. 30):

$$U_\mathrm{G} + \frac{U_\mathrm{P}}{2} \rightarrow 1 + i\sigma V_{\mathrm{G},z}(x,y) + \frac{i\sigma V_{\mathrm{P},z}(x,y)}{2} \approx$$
$$\approx \exp\left[i\left(\varphi_\mathrm{G}(x,y) + \frac{\varphi_\mathrm{P}(x,y)}{2}\right)\right] = \exp\left[i\varphi(x,y)\right]. \qquad (34)$$

The reconstruction consists of the following steps:

(1) Squared root of CBED distribution multiplied with the phase distribution of the reference wave $U_\mathrm{G}$: $\dfrac{U_\mathrm{G}}{\left|U_\mathrm{G}\right|}\sqrt{\mathrm{CBED}}$.

(2) The resulting wavefront is backward propagated to the sample plane by taking inverse FT of the result of (1).

(3) The phase distribution $\varphi(x,y)$ is extracted from the result of (2), and the phase distribution of the patch is calculated as $\varphi_P(x,y) = 2\left[\varphi(x,y) - \varphi_\mathrm{G}(x,y)\right]$.

As an example, a CBED pattern of a rotated graphene patch on graphene lattice, shown in Fig. 7a, was simulated and reconstructed. To mimic realistic experimental conditions, only CBED spots up to 5th order were considered in the reconstruction routine. The phase reconstructions obtained from the entire CBED pattern are shown in Fig. 7b − c. A slight background signal superimposed onto the reconstruction is due to additional terms caused by approximation in Eq. 33. When the positions of atoms are irregular, as in the patch shown in Fig. 7d, the reconstructed phase distribution exhibits atomic positions that are arranged into a regular lattice, not clear why but the reconstruction becomes very noisy, shown in Fig. 7e. When noise is added to CBED pattern the resulting reconstruction also becomes more noisy, but the atomic positions are retrieved at the same positions as from noise-free CBED pattern, shown in Fig. 7f. The recovered phase shift values are slightly below the original phase shift values (compare, for example, Fig. 7a and Fig. 7b and c), but it was noticed that when more higher-order CBED spots are included into the reconstruction, the recovered phase shift values approached the original values. It is therefore recommended for high-resolution reconstruction to acquire CBED pattern with higher-orders spots.

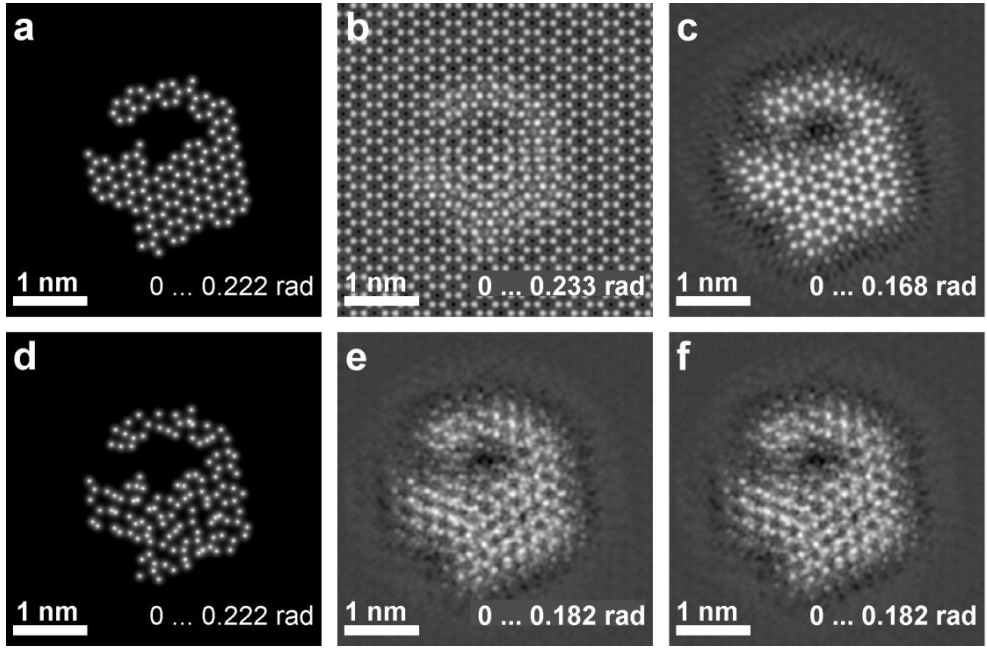

Fig. 7. Reconstruction obtained from an entire CBED pattern. (a) Phase distribution of the transmission function of the patch rotated by 10°. (b) Reconstructed phase distribution of the entire sample $\varphi(x, y)$. (c) Reconstructed phase distribution of the patch $\varphi_P(x, y)$. (d) Phase distribution of the transmission function of the patch rotated by 10°, where atoms are randomly shifted from their positions in the range 50 pm. (e) Reconstructed phase distribution $\varphi_P(x, y)$ from CBED pattern of (d). (f) Reconstructed phase distribution $\varphi_P(x, y)$ from CBED pattern of (d) where Gaussian noise was added to the CBED pattern such that signal-to-noise ratio = 5 for the first-order CBED spots.

## 2.4 Resolution

Lateral (in the $(x, y)$-plane) resolution obtained in reconstruction of an individual CBED spot can be evaluated from the Abbe criterion: $R = \dfrac{\lambda}{2NA}$, where $NA = \sin \alpha \approx \dfrac{D}{2\Delta f}$ and $D$ is the diameter of the imaged area. For the CBED pattern shown in Fig. 6 the resolution calculated from the parameters ($\Delta f = 2\,\mu\text{m}$, $D \approx 40\,\text{nm}$) amounts to 2.1 Å. This resolution also relates to the size of the probing beam in focus. The corresponding reconstructions obtained from individual CBED spots, shown in Fig. 6, exhibit the overall shape of the patch but the individual atoms cannot be resolved.

The lateral and axial (along the $z$-axis) resolution obtained in reconstruction from entire CBED pattern is given by $d_{x,y} = \dfrac{1}{\Delta k_{x,y}}$, and $d_z = \dfrac{1}{\Delta k_z}$, where $\Delta k_{x,y}$ and $\Delta k_z$ are the available range of $k$-values, $\Delta k_z = k - \sqrt{k^2 - k_{x,\max}^2 - k_{y,\max}^2}$. For the parameters used in the simulated example we calculate: $d_{x,y} = 0.3$ Å and $d_z = 1.9$ Å. It should be noted that these values are far from realistic values, because in the simulations we assumed a rather optimistic range of scattering angles detected up to 0.14 radian, when in reality scattering angles within only a few milliradians are detected. Also, for successful reconstruction, the supporting monolayer must provide a perfect reference signal, meaning that it should have no defects or ripples. Below we present experimental data which we reconstruct with the methods described just above and demonstrate the practical limits.

## 3. Experimental results

CBED patterns were acquired in a probe side aberration corrected Titan ChemiSTEM operated at 80 keV and a convergence semi angle, $\alpha$, of ~6 − 8 mrad. The convergence angle provides the expected lateral resolution of about $R = \dfrac{\lambda}{2NA} \approx 2.6 - 3.5$ Å when reconstructions are obtained from the zero-order CBED spot; it should be noted that the resolution does not depend on the defocus distance. During experiment the convergence angle was kept constant, and the sample height was changed by moving the sample along the optical axis. The images were acquired with a 16 bit intensity dynamic range detecting system.

The present study is not aimed to consider one material in particular, but periodical or hexagonal materials in general. The simulations were done for graphene and the experimental images were obtained on hBN. These two materials are only different in 1.8% of the lattice constant and slightly different scattering amplitudes. Since the chemical specificity of the scattering amplitudes was not accounted for in the simulation, the simulated images in the case of hBN would be the same as in the case of graphene, only re-scaled by 1.8%.

## 3.1 Defocus sequence

Acquired experimental CBED patterns of a hBN monolayer sample at different $\Delta f$ are shown in Fig. 8, where the diameter of the imaged area increases from 14 nm (at $\Delta f = 1\ \mu m$) to 84 nm (at $\Delta f = 6\ \mu m$). The hBN monolayer sample exhibits some occasional patches of adsorbates, as evident from the high angle annular dark field (HAADF) images shown in Fig. 8a – b. The HAADF image obtained after CBED pattern was acquired at $\Delta f = 6\ \mu m$ demonstrates that some patches vanished due to radiation damage by the beam (Fig. 8b).

Adsorbates and crystal deformations can be easily distinguished from each other by comparing the intensity distribution in the zero-order and higher-order CBED spots. For example, CBED pattern at $\Delta f = 2\ \mu m$ (Fig. 8d) exhibits an elongated dark feature in the first-order CBED spots that is not observed in the zero-order CBED spot. This feature can be attributed to a ripple between the two patches. Note that some elongated adsorbate structure is observed in the "before" HAADF image but it is not observed in the "after" HAADF image, Fig. 8a – b (indicated by the yellow arrow in Fig. 8b). This can imply that the structure was destroyed during the TEM studies but an imprint in form of a ripple remained in the hBN support. At the same time, a bright feature is observed in the zero-order CBED pattern at $\Delta f = 2\ \mu m$ (Fig. 8d) which can be attributed to a remaining fragment of the elongated structure.

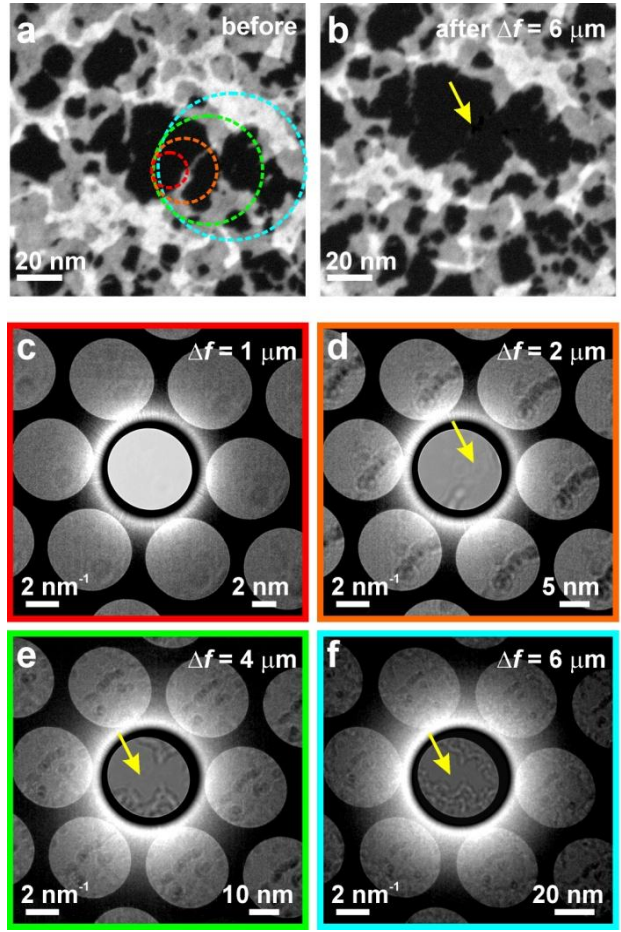

Fig. 8. Experimental CBED patterns of monolayer hBN at different sample height $\Delta f$. (a) –(b) High angle annular dark field (HAADF) images of the sample before and after CBED patterns acquisition. (c) – (f) CBED patterns at $\Delta f = 1, 2, 4$ and $6 \ \mu m$. The imaged area is marked by the corresponding colored circles in (a). The yellow arrow indicates the feature which is not observed in the zero-order spot but is observed in the first- and higher order CBED spots. The intensity of the central spot is reduced by factor $10^3$. The left scale bars correspond to the detector plane, the right scale bars correspond to the sample plane.

## 3.2 Underocus and overfocus imaging

Experimental CBED patterns of the same region of hBN sample acquired at $\Delta f = 5 \ \mu m$ and $\Delta f = -5 \ \mu m$ are shown in Fig. 9. HAADF images acquired before and after CBED imaging demonstrate no visible radiation damage of the sample (Fig. 9a − b). CBED patterns (Fig. 9c and e respectively) exhibit inversion of the contrast, as expected for weak phase objects. Also the intensity distributions within CBED spots exhibit centro-symmetrical flipping, see the magnified zero-order

CBED spots shown in Fig. 9b and c. These two effects are expected for $\Delta f < 0$ and $\Delta f > 0$ regimes as was discussed above.

In the CBED patterns shown in Fig. 9, the higher-order CBED spots show somewhat different intensity distributions than the zero-order spot. The explanation is that there are some surface perturbations such as ripples which can be imaged only in the first-order but not in the zero-order CBED spots. For example, for the CBED pattern shown in Fig. 9, there is a darker region in the first order CBED spots as indicated by magenta arrow. Such a dark region is not observed in the zero-order CBED spot and thus cannot be attributed to an adsorbate. The HAADF images of the sample also confirm that there are no adsorbates at this location, as indicated by the magenta arrows in Fig. 9a – b. Also, as already mentioned above, the intensity contrast due to atomic mis-arrangements is higher in the second-order than in the first-order CBED spots. Thus, the dark region can be explained by a surface deformation in that location, probably because of the stress between the two adjacent adsorbates.

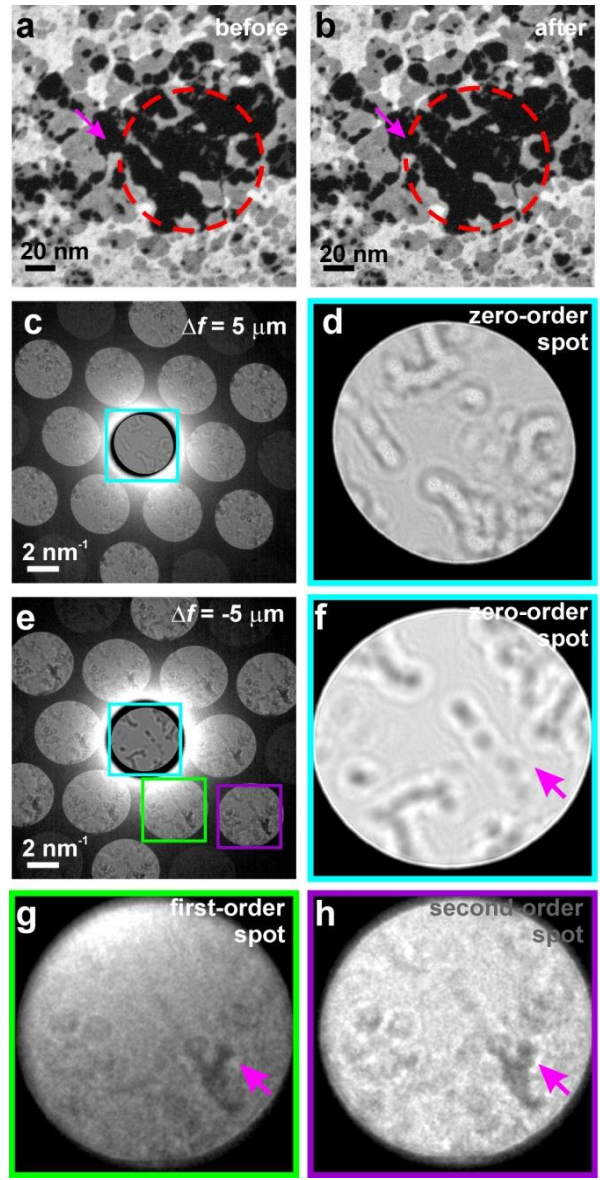

Fig. 9. CBED patterns of ML hBN acquired at $\Delta f > 0$ and $\Delta f < 0$.

(a) − (b) High angle annular dark field (HAADF) images of the sample before and after CBED imaging; the imaged area is marked by the red dashed circle.

(c) CBED pattern acquired at $\Delta f = 5$ μm and (d) the zero-order CBED spot distribution.

(e) CBED pattern acquired at $\Delta f = -5$ μm and magnified distributions of (f) the zero-order, (g) the first-order, and (h) the second-order CBED spots.

## 3.3 Reconstruction of zero-order CBED spot as an in-line hologram

Zero-order CBED spot of a CBED pattern can be treated as an in-line hologram and by calculating reconstruction integral as given by Eq. 29, the amplitude and phase distributions of the transmission function of the sample cab be obtained. Below we provide several examples of such reconstructions.

**Example 1.** HAADF images of the sample acquired before TEM studies are shown in Fig. 10a and b, there are no adsorbates in the studied area. The acquired CBED pattern (Fig. 10c) exhibits very weak intensity contrast in the zero-order CBED spot, shown in Fig. 10d. The amplitude and phase distributions reconstructed from the zero-order CBED spot are shown in Fig. 10e − g, they exhibit some adsorbate patches. One selected patch exhibits strong contrast in the amplitude distribution (Fig. 10f) which implies that the patch consists of some chemical elements that cannot be approximated as weak phase objects.

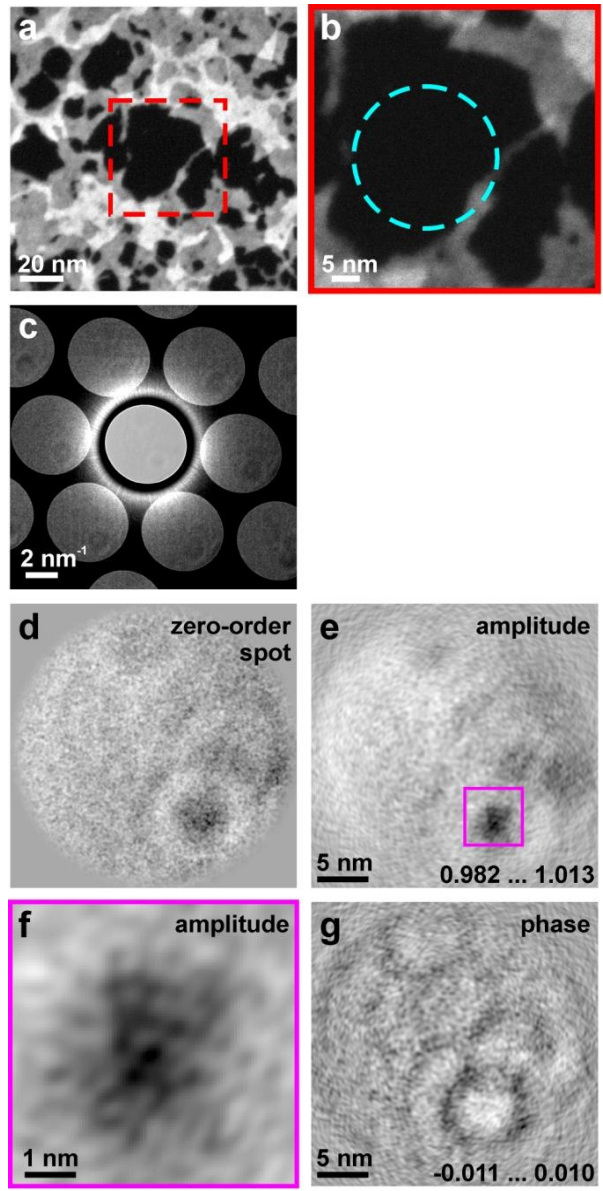

Fig. 10. Reconstruction of zero-order CBED spot as an in-line hologram. (a) and (b) high angle annular dark field (HAADF) images of the sample before CBED imaging. (c) CBED pattern. The imaged area is marked in (a) and (b). The intensity of the central spot is reduced by factor $10^3$. (d) The zero-order CBED spot and (e) − (f) the amplitude and (g) phase distributions reconstructed at the sample height $\Delta f = 2$ μm .

***Example 2***. Another example demonstrates the resolution limit of the reconstruction obtained from the zero-order CBED spot. HAADF image of the sample acquired before TEM studies (Fig. 11a − b) shows that there are some adsorbates in the studied area. The acquired CBED pattern exhibits strong intensity contrast in the zero-order CBED spot, shown in Fig. 11c − d. The reconstructed amplitude distribution, shown in Fig. 11e, exhibits blurred structure when compared to the reconstructed phase distribution shown in Fig.11f. This can be an indication that the object exhibits only phase-shifting properties and no absorption, so that its amplitude is close to one. For purely phase objects, the amplitude and phase distribution cannot be correctly recovered from a single reconstruction but they can be recovered by applying an iterative procedure[20, 21]. The reconstructed phase distribution, shown in Fig. 11f, shows the same patches distribution as imaged in the HAADF images. A selected feature (shown with the blue arrows in Fig. 11f) exhibits width of about 8 Å.

***Examples 3 and 4***. Two more examples of sample distributions reconstructed from the zero-order CBED spot are shown in Fig. 12 and 13. Adsorbate patches of sub-nanometre sizes are observed in the reconstructed phase distributions and can be cross-validated with the corresponding HAADF images. Also here, the reconstructed amplitude distributions exhibit blurred structure when compared to the reconstructed phase distributions, which can be an indication that the imaged objects exhibit only phase-shifting properties and their amplitude is close to one.

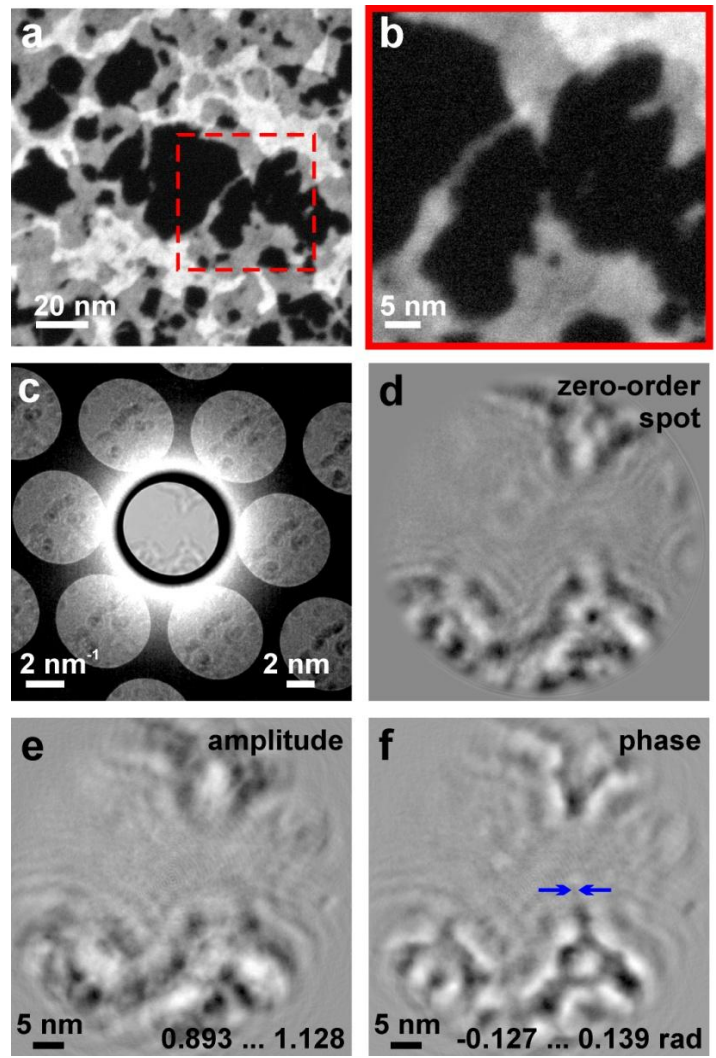

Fig. 11. Reconstruction of zero-order CBED spot as in-line hologram. (a) and (b) High angle annular dark field (HAADF) images of the sample before CBED patterns acquisition. (c) CBED pattern. The imaged area is marked in (a) and (b). The intensity of the central spot is reduced by factor $10^3$. (d) The zero-order CBED spot distribution and (e) and (f) the phase and amplitude distributions reconstructed at $\Delta f = 3.6 \ \mu\text{m}$.

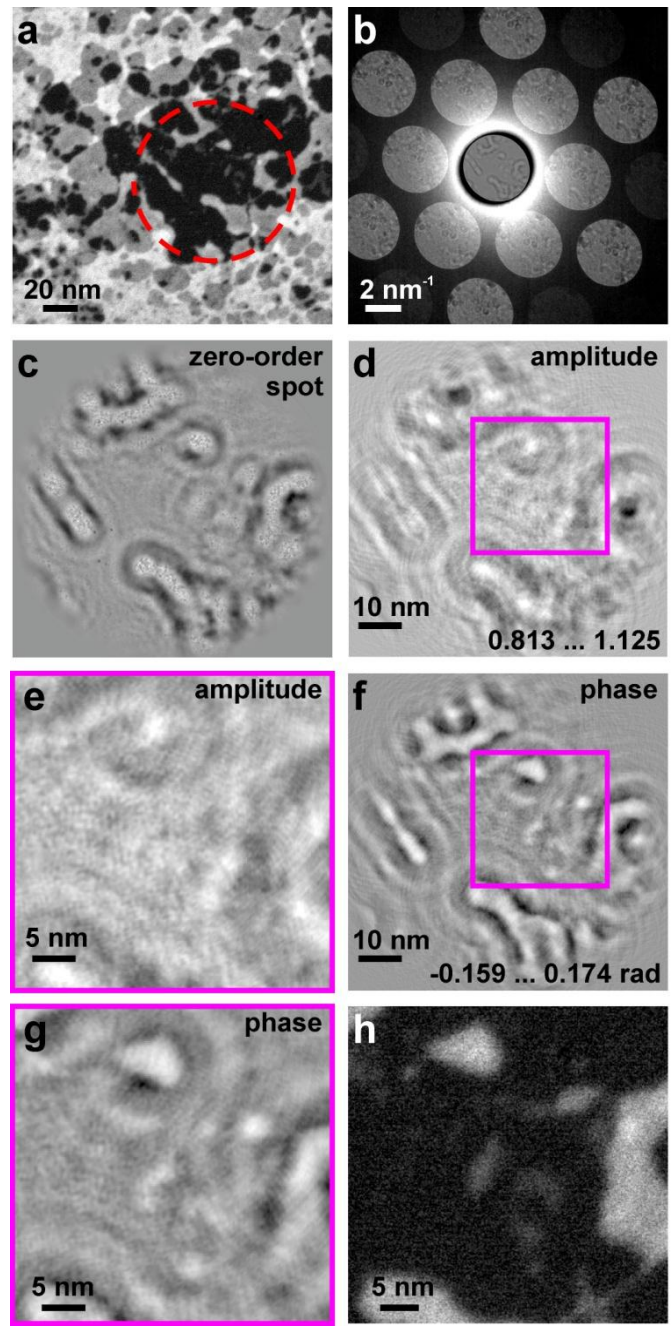

Fig. 12. Reconstruction of zero-order CBED spot as in-line hologram. (a) High angle annular dark field (HAADF) images of the sample before CBED imaging; the imaged area is marked by the red dashed circle. (b) CBED pattern. (c) Zero-order CBED spot. (d) – (e) The amplitude distribution reconstructed from the zero-order CBED spot at $\Delta f = 5\ \mu m$. (f) – (g) The phase distribution reconstructed from the zero-order CBED spot at $\Delta f = 5\ \mu m$. (h) Zoomed-in region in the HAADF image for comparison the reconstructed amplitude and phase distributions of the same region (marked by the magenta square).

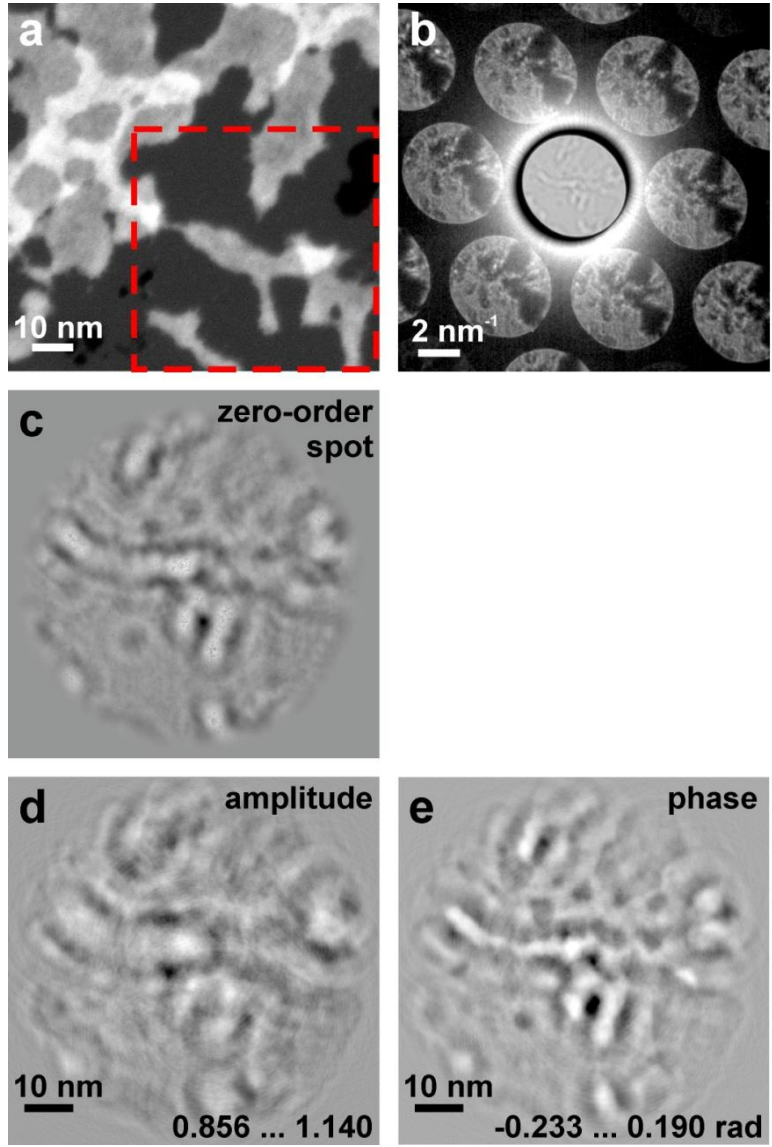

Fig. 13. Reconstruction of zero-order CBED spot as in-line hologram. (a) High angle annular dark field (HAADF) image of the sample area; the imaged area is marked by the red dashed square. (b) CBED pattern. (c) The zero-order CBED spot. (d) and (e) amplitude and phase distributions reconstructed from the zero-order CBED spot at $\Delta f = 5\ \mu m$.

## 3.4 Reconstruction of entire CBED pattern

In the reconstruction obtained from an entire CBED pattern the resolution of the reconstructed structures is given by the order of the available CBED spots. For example, if CBED pattern exhibits spots only up to the first order, no atomic resolution can be expected in the reconstruction. For such CBED patterns, reconstruction of only zero-order CBED spots is meaningful; this is the case for all CBED patterns shown in Fig. 8, 10, 11 and 13. For a successful high-resolution reconstruction the acquired CBED pattern should satisfy the following conditions: (1) It should include higher than the first-order CBED spots. (2) The range of measured *K*-values also defines the pixels size in the sample domain. Thus, the range of measured k-values must be selected to allow sampling with at least 2 pixels between the atoms. (3) The pixel size in the detector plane should be selected such that the reconstructed area size is larger than the illuminated area size. (4) The signal-to-noise ratio should allow detection of interference patterns in the higher-order CBED spots.

At the moment of the experimental acquisition we were not aware of these conditions, and we did not acquire CBED patterns with higher-order CBED spots. An attempt to obtained high-resolution reconstruction from CBED pattern which exhibit up to the second-order CBED spots is shown in Fig. 14. The reconstruction procedure was the same as described above where the reference wave was created as follows. A lattice with period that would match the position of CBED spots was created and the scattered wave was simulated. The measured CBED pattern was low-pass filtered (blurred) until no interference pattern was observed in the CBED spots. The phase of the simulated scattered wave was superimposed with the amplitude obtained from the low-pass filtered experimental CBED pattern, the resulting complex-valued distribution gave the reference wave in the detector plane. Although the resolution in the reconstruction shown in Fig. 14d nearly allows resolve individual atoms, the lattice deformations can be clearly seen in the reconstruction.

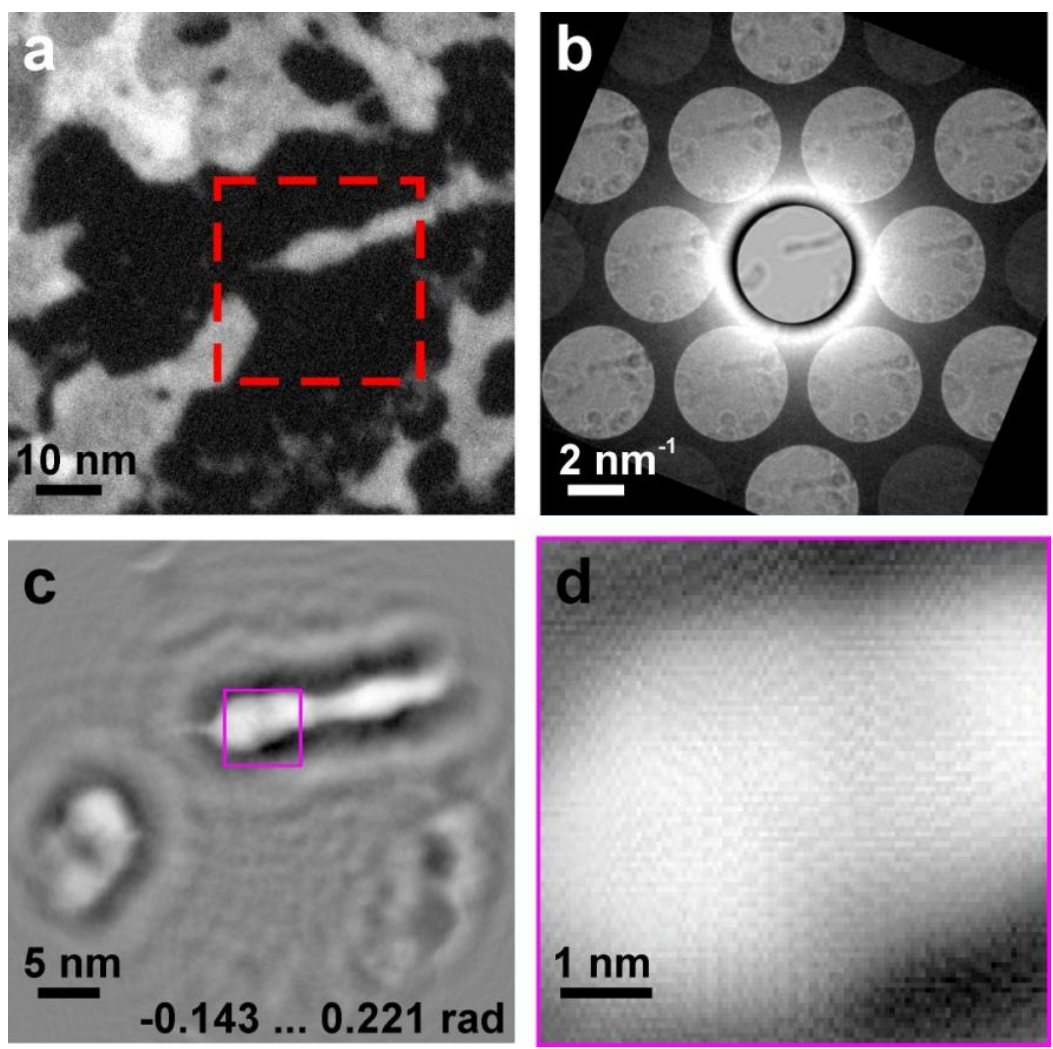

Fig. 14. Reconstruction of entire CBED spot. (a) High angle annular dark field (HAADF) image of the sample area; the imaged area is marked by the red dashed square. (b) CBED pattern. (c) Phase distribution reconstructed from the zero-order CBED spot by treating the CBED spot as an in-line hologram (low-resolution reconstruction). (d) Phase distribution of the adsorbate reconstructed from the entire CBED.

## 4. Discussion/Conclusion

In conclusion, we showed that CBED of two-dimensional monolayer crystals allows imaging adsorbates and lattice deformations in one single acquisition. Any shift of an atom from its ideal position in the lattice directly translates into enhanced or reduced intensity value. The symmetry of the intensity distribution can be traced back to whether the atomic shift occurred in lateral or in axial directions. By comparing the intensity distributions in zero- and higher order CBED spots, the contributions from adsorbates and lattice deformations can be clearly separated. When the probing wavefront is changed from convergent to divergent, weak phase adsorbates cause a contrast inversion in the CBED spots intensity distributions. It is known that in-line holography (also called defocus imaging) is the best approach to acquire weak phase shifting objects: weak phase objects create a high contrast interference pattern, while they exhibit no contrast when imaged in in-focus mode. A prominent example was imaging DNA molecules embedded into ice, reported by Matsumoto et al, where the in-focus image exhibited too weak contrast but presence of DNA molecules could be identified in their in-line holograms[22]. Each individual CBED spot can be treated as in-line holograms and the distribution of adsorbates can be reconstructed. The resolution of the reconstructed distributions can be evaluated by the Abbe criterion and provided by the divergence angle. The convergence angle can be increased until the CBED spots begin to overlap, which occurs at

$$\sin \alpha \approx \frac{1}{2} \sin \vartheta^{(1)}$$ where $\vartheta^{(1)}$ is given by the first diffraction order $\sin \vartheta^{(1)} = \frac{\lambda}{d^{(1)}}$, where $d^{(1)}$ is the

lattice period. This gives the maximal possible resolution of $R = \frac{\lambda}{2 \sin \alpha} \approx \frac{\lambda}{\sin \vartheta^{(1)}} = d^{(1)}$ which does not depend on the wavelength of the probing electrons. The obtained resolution of about 3 Å makes the CBED imaging potentially interesting for imaging individual biological macromolecules deposited onto two-dimensional crystal support[23]. The possibility to achieve high resolution in in-inline holographic imaging was recently reported by Adaniya et al, who imaged gold nano-particles and carbon nano-fibers by employing 20 keV energy electrons, at a resolution of ~1 nm defined by the divergence angle of 4.2 mrad[24]. The idea of studying biological macromolecules deposited onto two-dimensional crystal support has been intensively explored lately[25]. The reconstructions of polymersome deposited onto graphene were obtained from a sequence of 40 focal-series images[26] by the exit wave reconstruction method[27]. We showed that, in principle, a single CBED pattern is sufficient to reconstruct the sample distribution, which potentially allows minimizing the electron dose required for imaging individual biomolecules. We also showed in the simulated examples that reconstructions at atomic resolution are possible provided the acquired CBED pattern satisfies the following conditions: it contains the higher-order CBED spots, it is sampled with $\Delta k$ values that allow reconstruction of the area that exceeds the illuminated area, and the signal-to-noise ration in higher diffraction spots is sufficient to resolve the interference pattern. To conclude, CBED imaging allows encoding the exact three-dimensional atomic positions in CBED patterns, which then in principle can be accurately retrieved by applying a corresponding numerical procedure. $\Delta K$